\documentclass[Physsubmission, Phys]{SciPost}

\hypersetup{
    colorlinks,
    linkcolor={red!50!black},
    citecolor={blue!50!black},
    urlcolor={blue!80!black}
}

\usepackage[bitstream-charter]{mathdesign}
\DeclareSymbolFont{usualmathcal}{OMS}{cmsy}{m}{n}
\DeclareSymbolFontAlphabet{\mathcal}{usualmathcal}

\graphicspath{ {figs/} }

\begin{document}

\begin{center}{\Large \textbf{
      A semi-numerical method for one-scale problems applied to the
      $\overline{\rm MS}$-on-shell relation
}}\end{center}

\begin{center}
Matteo Fael\textsuperscript{1},
Fabian Lange\textsuperscript{1,2},
Kay Sch\"onwald\textsuperscript{1},
Matthias Steinhauser\textsuperscript{1$\star$}
\end{center}

\begin{center}
{\bf 1} Institut f{\"u}r Theoretische Teilchenphysik, Karlsruhe
Institute of Technology (KIT),\\ 76128 Karlsruhe, Germany
\\
{\bf 2} Institut f{\"u}r Astroteilchenphysik, Karlsruhe Institute of
Technology (KIT),\\ 76344 Eggenstein-Leopoldshafen, Germany
\\
* matthias.steinhauser@kit.edu
\end{center}

\begin{center}
\today
\end{center}


\definecolor{palegray}{gray}{0.95}
\begin{center}
\colorbox{palegray}{
  \begin{tabular}{rr}
  \begin{minipage}{0.1\textwidth}
    \includegraphics[width=35mm]{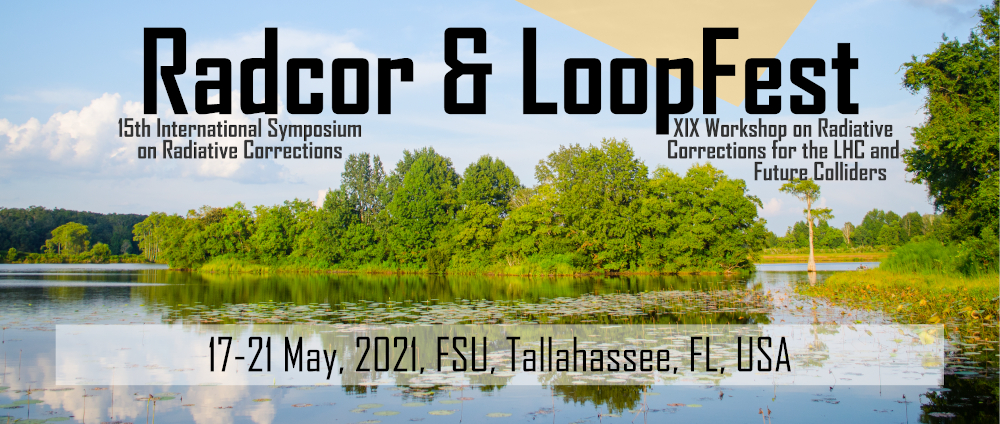}
  \end{minipage}
  &
  \begin{minipage}{0.85\textwidth}
    \begin{center}
    {\it 15th International Symposium on Radiative Corrections: \\Applications of Quantum Field Theory to Phenomenology,}\\
    {\it FSU, Tallahasse, FL, USA, 17-21 May 2021} \\
    \doi{10.21468/SciPostPhysProc.}\\
    \end{center}
  \end{minipage}
\end{tabular}
}
\end{center}

\section*{Abstract}
{\bf
  We discuss a practical approach to compute
  master integrals entering physical quantities
  which depend on one parameter.
  As an example we consider four-loop QCD corrections
  to the relation between a heavy quark mass
  defined in the $\overline{\rm MS}$ and on-shell
  scheme in the presence of a second heavy quark.
}


\section{Introduction}
\label{sec::intro}

At this conference a large number of very impressive and complicated
calculations to multi-loop and multi-leg processes have been
presented. 
With increasing number of scales and loops analytic results are more
and more difficult to obtain. It is thus often necessary to rely on
numerical approaches.  As an intermediate step we present in
this contribution a semi-analytic method, which we
discuss for a problem which depends on one parameter, $x$, usually the
ratio of two kinematic invariants.
Semi-analytic means that we construct numerical approximations for the
master integrals whereas all other steps of the calculation are
analytic.

In contrast to many other methods on the market which have a similar aim (see,
e.g.,
Refs.~\cite{Boughezal:2007ny,Lee:2017qql,Liu:2017jxz,Blumlein:2017dxp,Francesco:2019yqt,Hidding:2020ytt})
our approach is tailored to practical applications as has been demonstrated in
Ref.~\cite{Fael:2021kyg} where four-loop contributions to the
$\overline{\rm MS}$-on-shell relation with two mass scales have been computed.
We can treat systems which involve ${\cal O}(100)$ master integrals and obtain
a precision of the final result which is sufficient for phenomenological
applications.  At the moment our approach is formulated for problems which
depend only on one parameter. Furthermore, we do not aim for a precision of
hundreds of digits.

In these proceedings we review the findings of
Ref.~\cite{Fael:2021kyg} and provide further details on
the calculation. In particular we discuss results
for one non-trivial four-loop master integral.

\section{The method}
\label{sec::method}

The basic idea of our method is very simple: For a given set of master
integrals we establish the system of differential equations. Then we
compute the master integrals for a convenient value of $x=x_0$, which
is not necessarily physical, and use these results as boundary
conditions to construct a power-log expansion. Up to this point the
calculation is in general analytic. Let us assume we want to compute
the master integrals at the point $x=x_1$.  This is achieved with the
help of a power-log expansion around $x=x_1$, which is again
constructed with the help of the differential equations.  The boundary
conditions are obtained at a suitable value of $x=x_{01}$ between
$x_0$ and $x_1$, where both expansions converge, by evaluating
numerically the first expansion around $x=x_0$.  We call this step
{\it numerical matching}.  This step can be repeated, if necessary
several times, in order to arrive at any desired value of $x$.

There are only few requirements, which have to be fulfilled to apply this
method. In particular, it is not necessary to derive a system of differential
equations in Fuchsian or even canonical form.  One only has to avoid that
$1/\epsilon$ poles are present on the diagonal of the matrix obtained from the
differential equations. Furthermore, it is not necessary that the set of
master integrals is minimal. It is advantageous that the boundary conditions at
$x=x_0$ are analytic, however, in principle even they be can available in
numeric form. 
The major part of the CPU time is needed for deriving the linear system
of equations fulfilled by
the series coefficients of the master integrals, solving such linear
system and performing the
numerical matching.
In case one only has to deal with a simple Taylor
expansion it only takes several hours even in case a few dozens of expansion
terms are considered. On the other hand, in case one has a power-log expansion
the complexity increases significantly, in particular if two powers of
logarithms appear for each new order of $\epsilon$.

In the next Section we discuss eight colour factors for the
$\overline{\rm MS}$-on-shell relation. We consider four-loop two-scale
integrals with $x=m_2/m_1$ where for the external momentum $q$ we have
$q^2=m_1^2$. The mass $m_2$ appears in closed fermion loops.  In the
limit $m_2\gg m_1$ an analytic evaluation is possible since one ends up
with tadpole integrals up to four loops and on-shell integral up to
three loops. For these kind of one-scale integrals both the reduction
to master integrals and analytic expressions of the latter are well
known. We compute 50 terms for the power-log expansion around $1/x=0$,
which shows good convergence properties, even down to $x\approx
1.5$. At this point we match to the Taylor expansion around $x=1$. The
expansion around $x=0$ is again a power-log expansion which we match
at $x=0.5$ to the $x=1$ expansion.

For our application it is indeed sufficient to construct only three
expansions.  In general it might be that the differential equations
contain poles which limit the radius of convergence. In these cases
on has to introduce further matching points. In case these poles are
spurious a simple Taylor expansion is in general sufficient. For the cases where
the poles have a physical origin it might be that a power-log
expansion is necessary.

\section{Mass relation}
\label{sec::MSOS}

\begin{figure}[t]
  \centering
    \begin{tabular}{cccc}
      \includegraphics[width=0.2\textwidth]{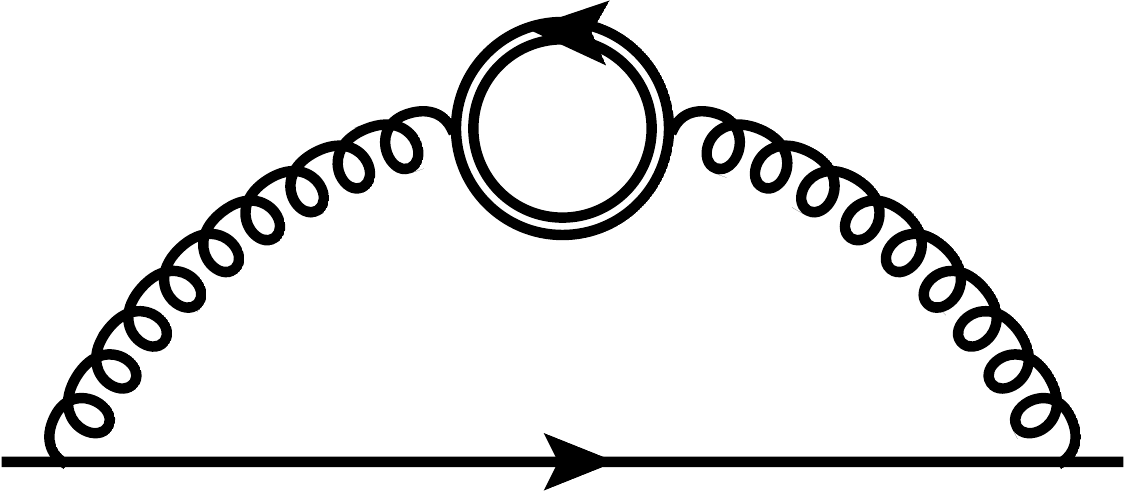} &
      \includegraphics[width=0.2\textwidth]{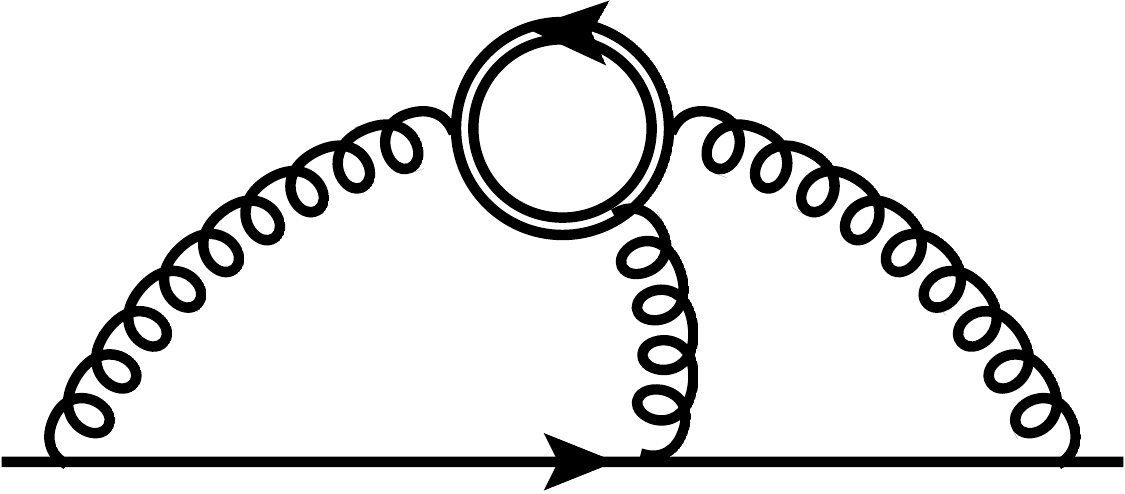} &
      \includegraphics[width=0.2\textwidth]{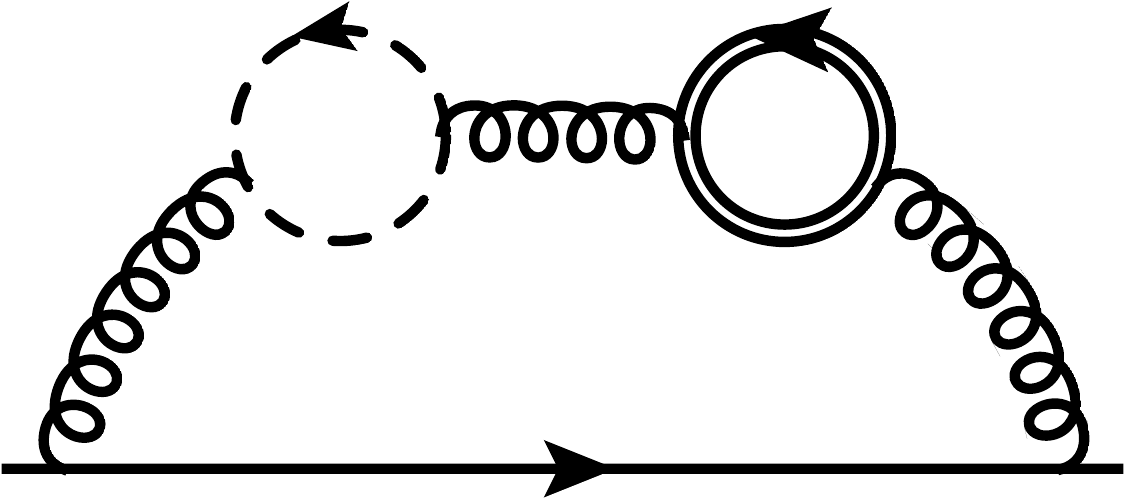} &
      \includegraphics[width=0.2\textwidth]{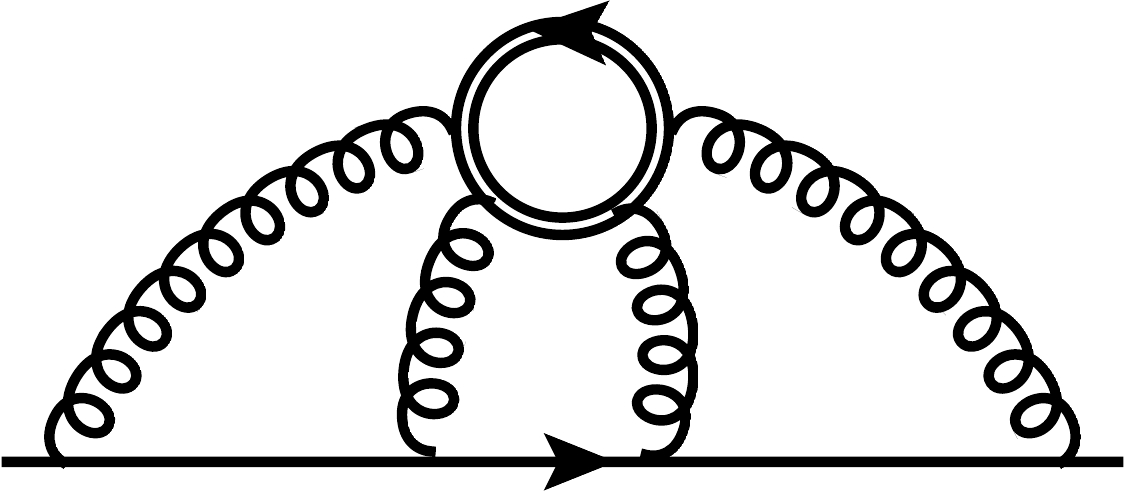} \\
      (a) & (b) & (c) & (d) \\
      \includegraphics[width=0.2\textwidth]{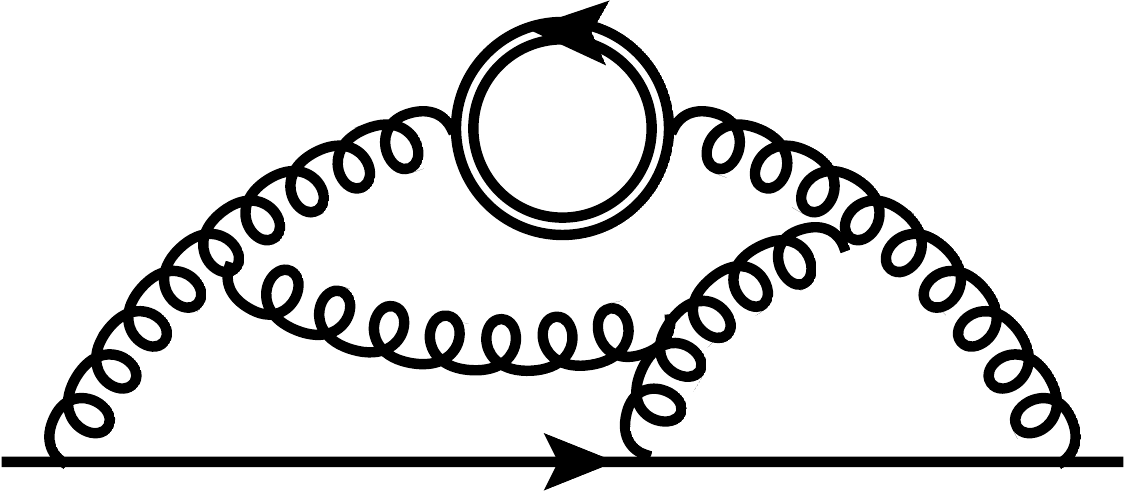} &
      \includegraphics[width=0.2\textwidth]{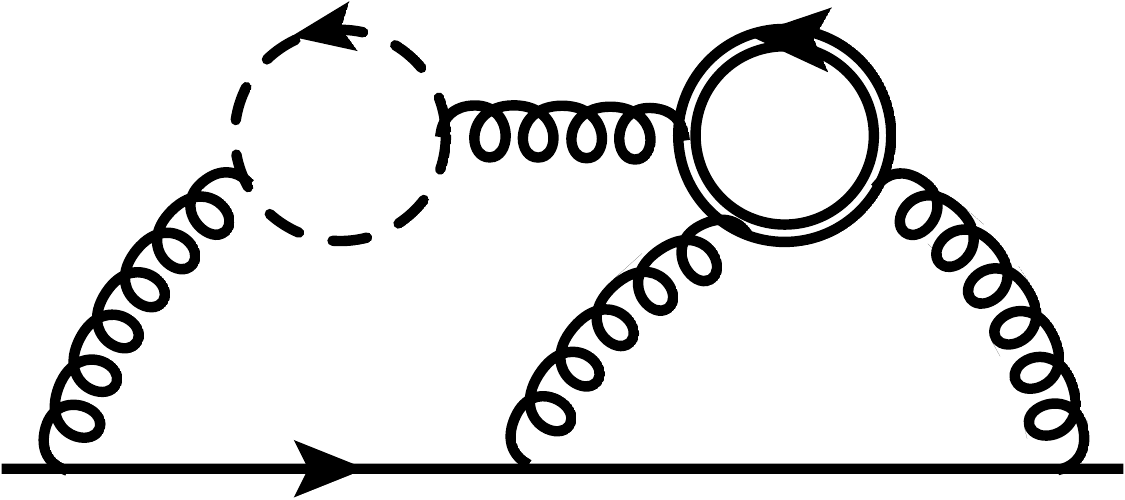} &
      \includegraphics[width=0.2\textwidth]{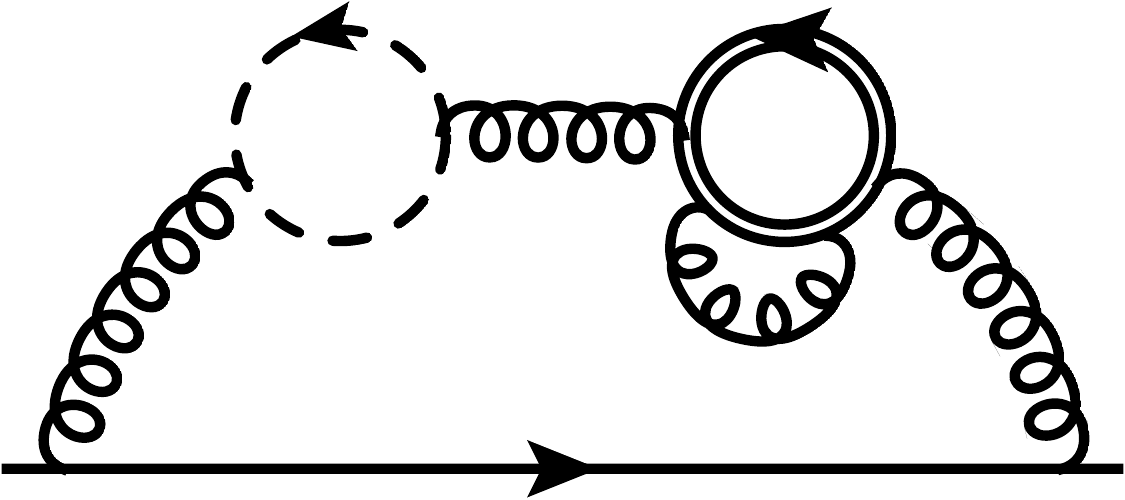} &
      \includegraphics[width=0.2\textwidth]{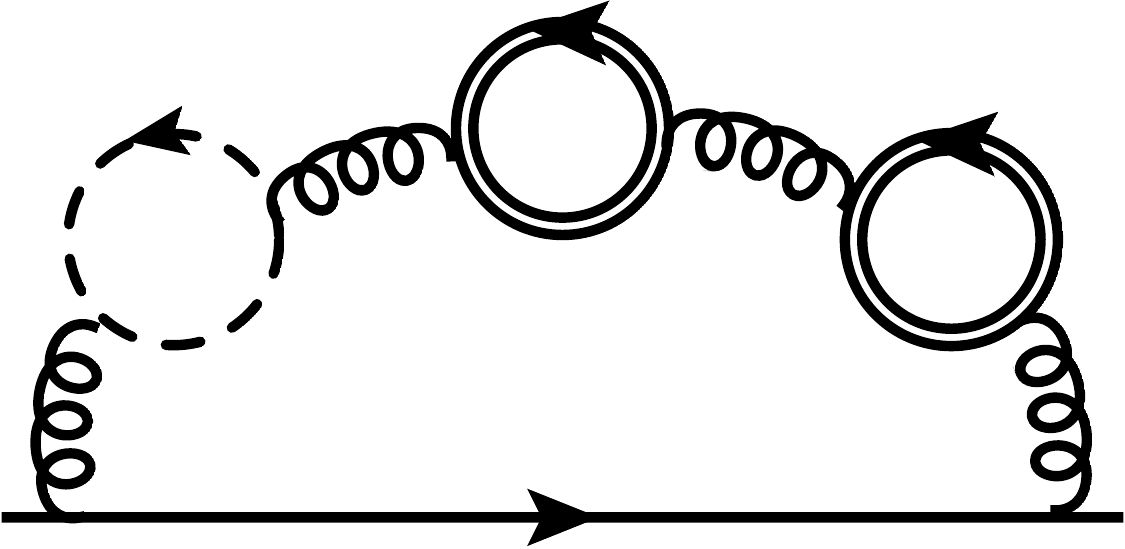} \\
      (e) & (f) & (g) & (h) \\
    \end{tabular}
  \caption{\label{fig::FDs}Sample Feynman diagrams contributing to
    relation between the $\overline{\rm MS}$ and on-shell
    mass. Straight and curly lines represent quarks and gluons,
    respectively. Dashed and double lines represent massless fermions
    and fermions with mass $m_2$, respectively.}
\end{figure}

The quark mass relation, which is considered in this section, involves
renormalized heavy quarks in the most important renormalization schemes: the
on-shell and the $\overline{\rm MS}$ scheme. In
Ref.~\cite{Marquard:2015qpa,Marquard:2016dcn} it has been computed to
four-loop order assuming that all lighter quarks have mass zero. Analytic
results for the contributions involving two different masses are only known
at three loops from Refs.~\cite{Bekavac:2007tk,Fael:2020bgs}. At four loops such
corrections have been considered for the first time in
Ref.~\cite{Fael:2021kyg}. Results for the eight colour factors\footnote{There are 16
  colour factors in total.}
\begin{eqnarray}
  &&C_F T_F^3 n_m n_l^2,\quad C_F T_F^3 n_m n_l n_h,\quad 
    C_F T_F^3 n_m n_h^2,\quad C_F T_F^3 n_m^2 n_l,   \nonumber\\
  &&C_F T_F^3 n_m^2 n_h,\quad C_F T_F^3 n_m^3,\quad       
    C_F^2 T_F^2 n_m n_l,\quad C_FC_A T_F^2 n_m n_l,
     \label{eq::colour}
\end{eqnarray}
have been obtained using the method described in the previous Section.
They all either contain three closed fermion
loops or two closed fermion loops where one of them has the mass $m_2$
and the other is massless.  Sample Feynman diagrams are shown in
Fig.~\ref{fig::FDs}. The numerical results, which are available for
any value of $x=m_2/m_1$, have been compared to the analytic
expressions, which have also been obtained in~\cite{Fael:2021kyg}.

\begin{figure}[t]
  \centering
  \includegraphics[width=0.5\textwidth]{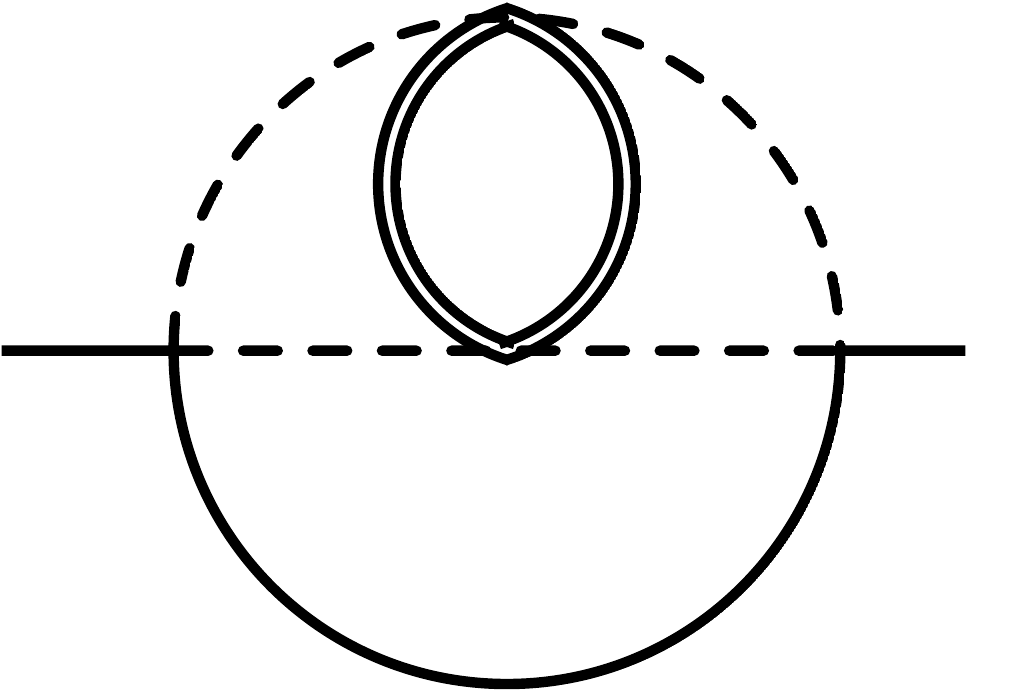}
  \caption{\label{fig::d4L456_7mv} Graphical representation of the master
    integral in Eq.~(\ref{eq::J}). Dashed, solid and double lines represent
    scalar propagators of mass 0, $m_1$ and $m_2$.}
\end{figure}

For our calculation we have to consider 339 master integrals.
In the following we discuss in detail the seven-line master integral
(see Fig.~\ref{fig::d4L456_7mv})
\begin{eqnarray}
  J(x) &=& \verb|d4L456[1, 0, 0, 1, 1, 1, 1, 0, 0, 1, 1, 0, 0, 0]|\,,
 \label{eq::J}
\end{eqnarray}
which starts at order $1/\epsilon^4$. In our calculation it is needed
including the ${\cal O}(\epsilon)$ terms.  In Ref.~\cite{Fael:2021kyg} the
expansion around $x=0$, $1$ and $\infty$ have been computed using the
approach discussed in the previous Section.  Furthermore, the analytic
result could be obtained.  Note, however, that the ${\cal
  O}(\epsilon)$ term of $J(x)$ contains cyclotomic harmonic
polylogarithms
up to
weight 6, which cancel in the proper sum for the quark mass relation.
It is  both a non-trivial task to compute them numerically and to
perform an analytic expansion. However, it is straightforward to
obtain its value at $x=0$:
\begin{eqnarray}
  J(0) &=&
  - \frac{1}{12\epsilon^4}
  - \frac{13}{24 \epsilon^3}
  + \frac{1}{\epsilon^2} \left(-\frac{15}{16} - \frac{13\pi^2}{36}\right)
  + \frac{1}{\epsilon}    \left( \frac{1135}{96} - \frac{169\pi^2}{72}
  - \frac{86\zeta_3}{9}\right)
  \nonumber\\&&
  + \frac{28699}{192}
  - \frac{65\pi^2}{16}
  - \frac{559\zeta_3}{9}
  - \frac{149\pi^4}{90}
  + \epsilon \left(
  \frac{144429}{128}
  + \frac{14755\pi^2}{288}
  - \frac{227\zeta_3}{2}
  \right.\nonumber\\&&\left.
  - \frac{1937\pi^4}{180}
  - \frac{1118 \pi^2 \zeta_3}{27}
  - \frac{7604 \zeta_5}{15}
  \right)
  + {\cal O}(\epsilon^2)
                       \,.
  \label{eq::J0}
\end{eqnarray}
For this reason we perform in the following the comparison with the
exact results only for $x=0$. Note that our starting point is 
the $x\to\infty$ limit.

\begin{figure}[t]
  \centering
  \includegraphics[width=0.9\textwidth]{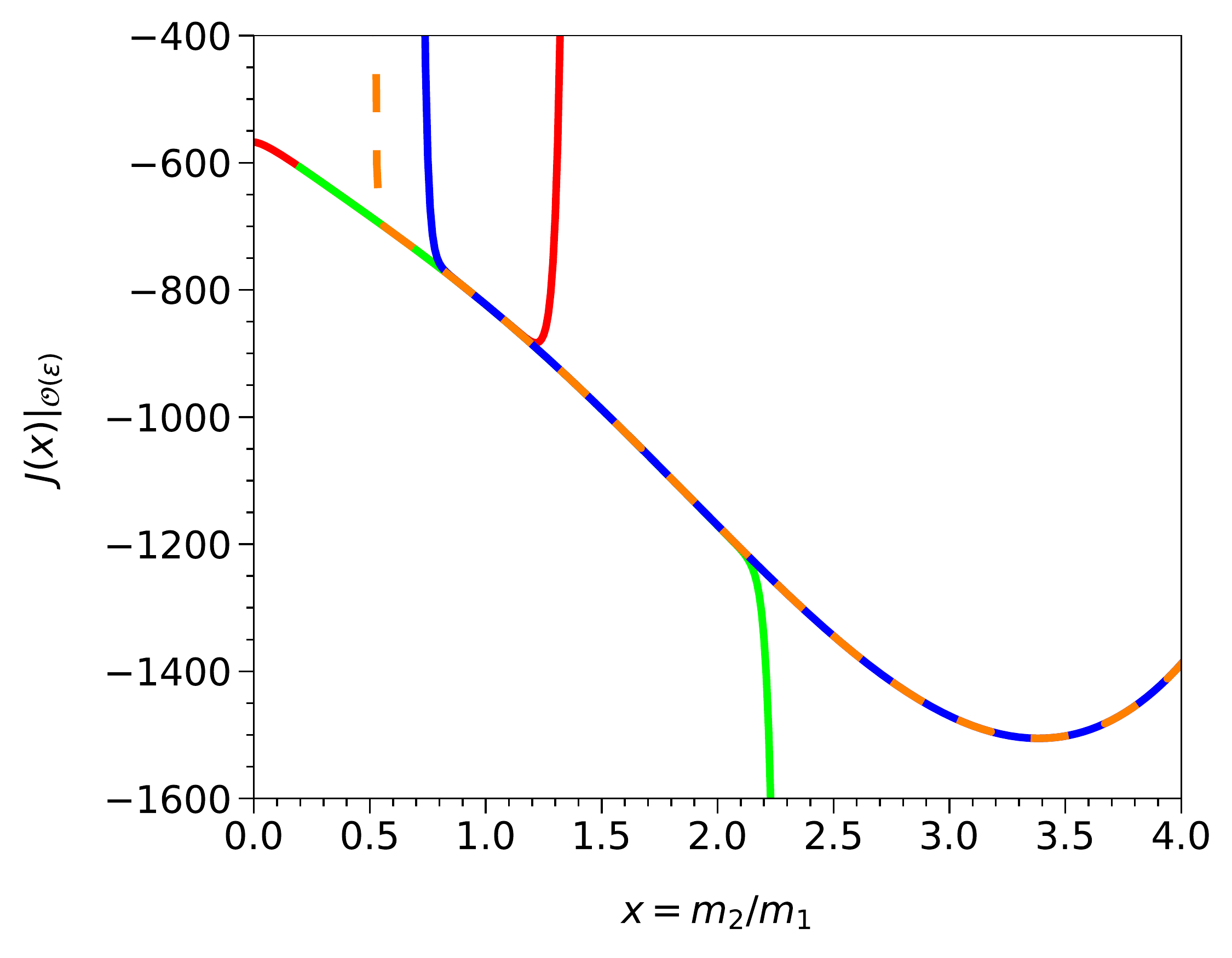}
  \caption{\label{fig::J}
    ${\cal O}(\epsilon)$ term of the integral $J$.
    The solid red, green and blue curves correspond to the expansions
    in $x$, $1-x$ and $1/x$. For the dashed orange curve the expansion
    parameter $1-1/x$ has been used.}
\end{figure}

In Fig.~\ref{fig::J} we show results for the ${\cal O}(\epsilon)$ term of
$J(x)$ for $0\le x \le 4$ where the red, green and blue solid curves correspond to
the expansions in $x$, $1-x$ and $1/x$, including 50 expansion terms. They are
the immediate results of our method and perfectly cover the whole $x$ range.
The dashed orange curve corresponds to the 
expansion in $1-1/x$ which is obtained from the $1-x$ expansion. 
It is interesting to note that the green curve leads to
better results for $x<1$ whereas $(1-1/x)$ is the better choice for the
expansion parameter for $x>1$. In fact, in the plot there is no visible
difference between the blue and orange curve.
Even for x=8 the deviation is still below 10\%.

\begin{figure}[t]
  \centering
  \includegraphics[width=0.9\textwidth]{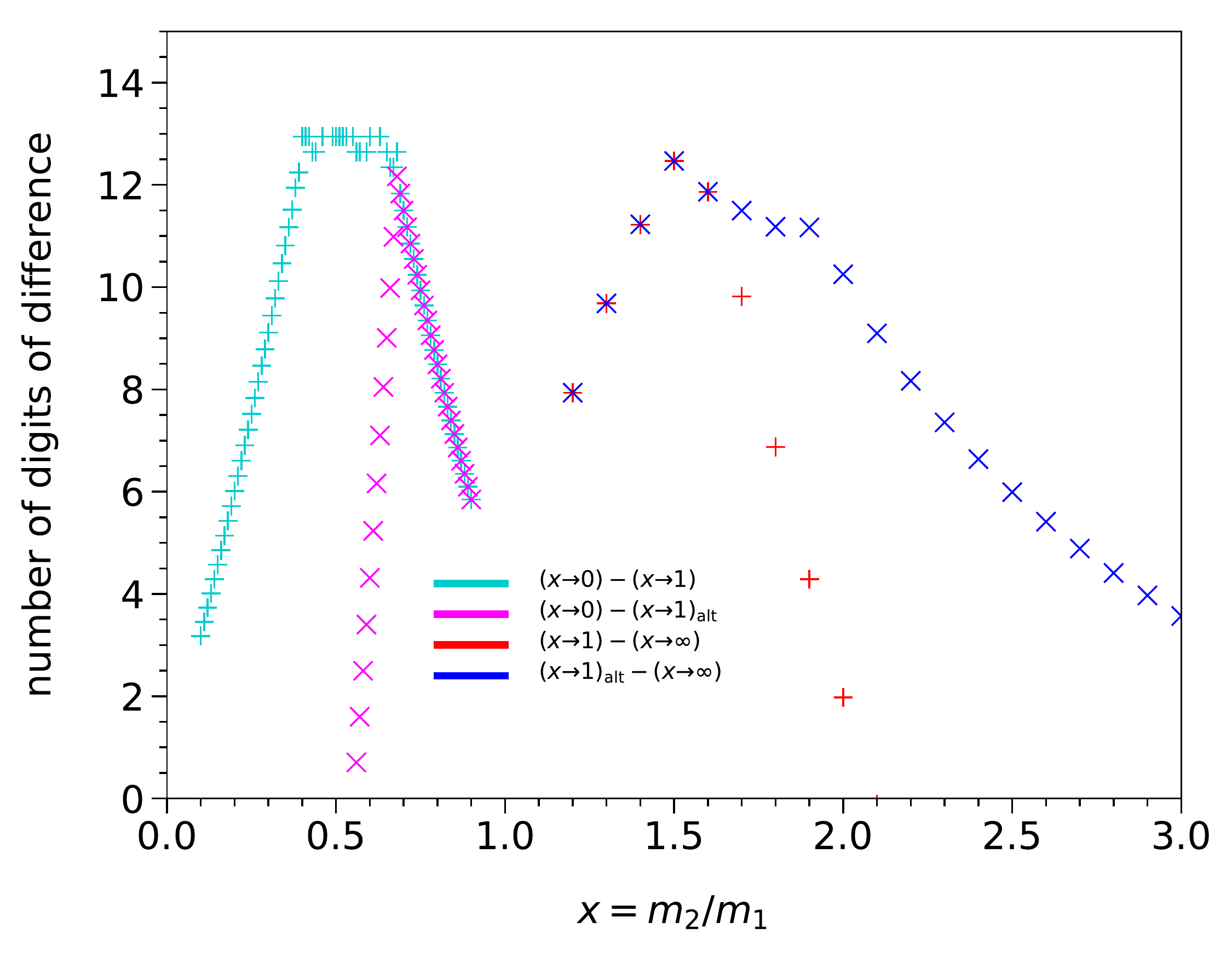}
  \caption{\label{fig::MI456} Difference of expansions. The number of
    digits on the $y$ axis is given by ``$-\log_{10}(\mbox{difference})$''.}
\end{figure}

Let us next discuss the numerical accuracy. We remark that we can reproduce
the analytic result for $x=0$ in Eq.~(\ref{eq::J0}) with a relative precision
of $10^{-13}$ for the ${\cal O}(\epsilon)$ term.  The lower $\epsilon$ terms
are even more precise; for example, the ${\cal O}(\epsilon^0)$ term has a
precision of $10^{-17}$. Remember that $J(0)$ is obtained from analytic
calculations at $x\to \infty$ and numerical matching for $x=0.5$ and $x=1.5$.
In order to quantify the quality of the approximations we consider in
Fig.~\ref{fig::MI456} differences
of expansions of the ${\cal O}(\epsilon)$ term.  The
interesting regions are around the matching points $x=0.5$ and $x=1.5$ where
we observe differences of order $10^{-12}$ to $10^{-13}$.  Taking into account
that the ${\cal O}(\epsilon)$ term itself is of order $10^2-10^3$
(cf. Fig.~\ref{fig::J}) we can claim a relative precision of more than
14 digits. Away from the matching points the respective expansion
provides even better results.

In Fig.~\ref{fig::MI456} we show two versions of the expansion around
$x=1$. The cyan and red curves use $(1-x)$ as expansion parameter and the
pink and blue curves use $(1-1/x)$. As already observed in Fig.~\ref{fig::J}
we again deduce that $(1-1/x)$ is a better
choice for $x>1$ whereas $(1-x)$ is better suited from $x<1$.

In Ref.~\cite{Fael:2021kyg} expansions for all 339 master integrals have been
obtained and then the mass relations for the eight colour factors of
Eq.~(\ref{eq::colour}) have been constructed. In Ref.~\cite{Fael:2021kyg} also
(exact) analytic results for all 339 master integrals have been computed. This
allows for a comparison of the final physical quantity. We could show that the
agreement between the exact and approximated results is at least 10
significant digits for the bare four-loop quantity. For most colour factors
this is also true for the renormalized quantities. Due to logarithmic
divergences (e.g., the colour structure $C_F^2n_ln_h$ for $x\to0$) one
observes strong cancellations in some regions of the parameters
space. However, we still have 5~significant digits in the final expression.
For more details we refer to Ref.~\cite{Fael:2021kyg}.
This is promising for the remaining eight colour factors where most likely no
analytic results in terms of iterated integrals can be obtained.

\section{Conclusion}

In this contribution we have described a method which can be used to
obtain semi-analytic results for multi-loop problems which depend on
one parameter $x$.  As an example four-loop corrections to the heavy
quark mass relation with two mass scales, $m_1$ and $m_2$, have been
considered. Using analytic results for $m_2\gg m_1$ it is possible to
use the differential equations for the master integrals
to transfer the information to $x=0$. The comparison to the
known analytic result shows that a precision between 5 and 10 digits
can be obtained. Using deeper expansions and/or further intermediate
matching points even a higher accuracy can be achieved.
Detailed result for one of the master integrals has been presented in this
contribution. For the results of the mass relation we refer to Ref.~\cite{Fael:2021kyg}.

\section*{Acknowledgements}
This research was supported by the Deutsche Forschungsgemeinschaft (DFG,
German Research Foundation) under grant 396021762 --- TRR 257 ``Particle
Physics Phenomenology after the Higgs Discovery''.




\bibliography{SciPost_Example_BiBTeX_File.bib}

\nolinenumbers

\end{document}